\documentclass[journal]{IEEEtran}
\usepackage{amsmath,amsfonts}
\usepackage{algorithmic}
\usepackage{algorithm}
\usepackage{array}
\usepackage[caption=false,font=normalsize,labelfont=sf,textfont=sf]{subfig}

\usepackage{hyperref}
\usepackage{textcomp}
\usepackage{stfloats}
\usepackage{url}
\usepackage{verbatim}
\usepackage{graphicx}
\usepackage{cite}
\usepackage{multirow}
\usepackage{multicol}
\usepackage{enumitem}
\usepackage{pifont}
\usepackage{color}
\usepackage{makecell}
\usepackage{subfig}

\begin{document}

\title{AdapCsiNet: Environment-Adaptive CSI Feedback via Scene Graph-Aided Deep Learning}

\author{Jiayi~Liu,
~Jiajia~Guo,
\IEEEmembership{\normalsize
{Member,~IEEE}},
~Yiming~Cui,
~Chao-Kai~Wen, 
\IEEEmembership{\normalsize
{Fellow,~IEEE}},
and Shi~Jin, 
\IEEEmembership{\normalsize
{Fellow,~IEEE}}

}		

\maketitle
 
\begin{abstract}
Accurate channel state information (CSI) is critical for realizing the full potential of multiple-antenna wireless communication systems. While deep learning (DL)-based CSI feedback methods have shown promise in reducing feedback overhead, their generalization capability across varying propagation environments remains limited due to their data-driven nature. Existing solutions based on online training improve adaptability but impose significant overhead in terms of data collection and computational resources. In this work, we propose AdapCsiNet, an environment-adaptive DL-based CSI feedback framework that eliminates the need for online training. By integrating environmental information---represented as a scene graph---into a hypernetwork-guided CSI reconstruction process, AdapCsiNet dynamically adapts to diverse channel conditions. A two-step training strategy is introduced to ensure baseline reconstruction performance and effective environment-aware adaptation. Simulation results demonstrate that AdapCsiNet achieves up to 46.4\% improvement in CSI reconstruction accuracy and matches the performance of online learning methods without incurring additional runtime overhead.

\end{abstract}

\vspace{-0.6cm}
\section{Introduction} 

Multiple-antenna technology has significantly advanced wireless communications, enabling substantial improvements in spectral and energy efficiency \cite{andrews20246}. To fully exploit these benefits, accurate downlink channel state information (CSI) is essential. However, achieving high CSI accuracy comes at the cost of substantial feedback overhead, necessitating the development of efficient CSI feedback techniques \cite{lar2014next}. 

In recent years, deep learning (DL)-based methods have demonstrated significant potential in CSI feedback, particularly in reducing feedback overhead while maintaining high reconstruction accuracy. The first DL-based CSI feedback framework, CsiNet \cite{wen2018deep}, employs an autoencoder to learn the mapping function for compressing and recovering high-dimensional CSI. Following this work, various DL-based approaches have been explored to enhance CSI feedback \cite{guo2022overview}, including the design of novel neural network (NN) architectures \cite{9926175} and the integration of multi-domain correlations \cite{wang2021compressive}.

Despite achieving notable performance gains, DL-based CSI feedback suffers from a key limitation: insufficient generalization capability. This issue arises from its data-driven nature \cite{10802972}. When the propagation environment undergoes significant changes, the underlying channel distribution also shifts, causing a mismatch between the training and testing data. Consequently, CSI feedback NNs experience severe performance degradation \cite{baraniuk2020science}. To mitigate this issue, existing studies have explored online training techniques to improve the environment adaptability of feedback NNs \cite{guo2022overview}. The authors in \cite{wang2021multi,10622316,online2023csi} propose fine-tuning pre-trained feedback NNs with a small number of CSI samples from new environments, effectively minimizing data requirements.
The study \cite{adap2024smalldataset} proposes a scenario-adaptive plug-in module for CSI translation, requiring only this module to be trained when encountering unfamiliar environments.
Furthermore, the studies in \cite{10622316,ContinuousOL2024CSI} employ continuous online learning to mitigate catastrophic forgetting during online training. 
Despite considerable efforts, these methods still demand extensive CSI collection and training,  making them infeasible for practical implementation.

Another approach to enhance the environment adaptability of feedback NNs is the model switch mechanism, which involves training multiple environment-specific feedback NNs to be selected based on the environment \cite{zhang2023switch,bathala2023multi,enviraware2023CSI,zhang2024zone}.
The frameworks proposed in \cite{zhang2023switch,bathala2023multi} independently train indoor- and outdoor-specific CSI reconstruction NNs, enabling the system to select either the indoor-trained or outdoor-trained decoder during inference based on the current test environment.
The study \cite{enviraware2023CSI}, aimed at enabling automated environment recognition, intelligently clusters CSI data with similar channel characteristics, while the study \cite{zhang2024zone} directly selects models based on the user equipment (UE) position.
Despite performance improvements, the model switch relies solely on basic environmental indicator (e.g., indoor or outdoor) and cannot effectively handle and adapt to diverse environments.

Since wireless channel characteristics are closely correlated with and determined by the propagation environment, environmental information can serve as valuable prior knowledge for characterizing the channel distribution. For instance, the authors in \cite{alkhateeb2023real} propose that various wireless channel parameters, such as propagation path characteristics and channel covariance information, can be inferred from a digital twin-modeled propagation environment. If such environmental information is integrated into DL-based CSI feedback, the NNs can gain awareness of underlying propagation conditions, significantly enhancing their adaptability to channel distribution variations.

In this work, we propose {\bf AdapCsiNet}, an environment-adaptive DL-based CSI feedback framework that eliminates the need for online training. By incorporating environmental information---represented as a scene graph---into the CSI reconstruction NN, AdapCsiNet enhances adaptability across diverse environments. The framework adopts a one-sided CSI feedback approach: UE compresses CSI into low-dimensional codewords using a fixed Gaussian projection matrix, and the base station (BS) reconstructs the CSI using a stack of neural layers. To enable environmental adaptability, the channel distribution of various propagation environments, represented by scene graphs, is integrated into the reconstruction process through a hypernetwork. This hypernetwork dynamically generates the parameters of the reconstruction NN based on the scene graph input.

To effectively address the challenge of adapting to varying environments, AdapCsiNet employs a two-step training strategy. First, the CSI reconstruction NN is pre-trained without scene graph integration to establish a robust baseline for general-purpose CSI reconstruction. Then, the hypernetwork is trained to refine this process by producing environment-specific parameters conditioned on the scene graph. Simulation results demonstrate that AdapCsiNet achieves effective adaptation to diverse environments, improving CSI reconstruction accuracy by up to 46.4\% and achieving performance comparable to online training methods---without incurring additional computational overhead.

\section{System model and CSI feedback}
\subsection{Signal Transmission Model}
\begin{figure*}[!t]
\centering
{\includegraphics[width=0.83\textwidth]{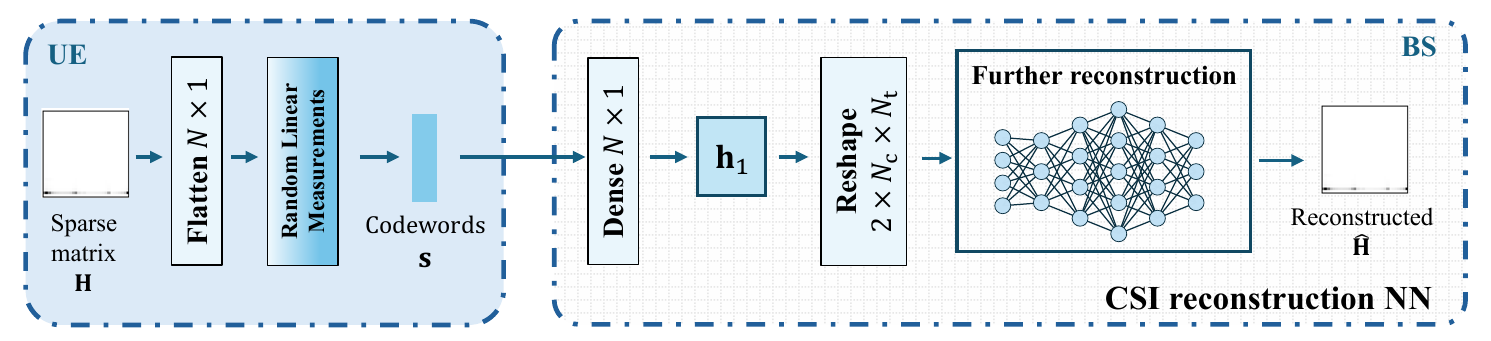}%
\caption{One-sided CSI reconstruction framework.}
\label{fig:framea}}
    \vspace{-0.6cm}
\end{figure*}
We consider a downlink transmission link equipped with multiple antennas. The BS is equipped with a uniform linear array with $ N_{\mathrm{t}} > 1 $ antennas, while the UE is equipped with a single antenna. The system employs orthogonal frequency division multiplexing (OFDM) with $ N'_{\mathrm{c}} $ subcarriers. The signal received at the $ n $-th subcarrier can be expressed as:
\begin{equation} \label{eq:y=hx+n}
y_n = \tilde{\mathbf{h}}_n^H \mathbf{v}_n x_n + z_n, \quad n = 1, 2, \dots, N'_{\mathrm{c}},
\end{equation}
where $ \mathbf{v}_n \in \mathbb{C}^{N_{\mathrm{t}} \times 1} $ is the precoding vector, $ x_n \in \mathbb{C} $ is the transmitted data symbol, and $ z_n \in \mathbb{C} $ represents the additive white Gaussian noise (AWGN). $ \tilde{\mathbf{h}}_n \in \mathbb{C}^{N_{\mathrm{t}} \times 1} $ denotes the channel vector on the $ n $-th subcarrier.  Consequently, the complete downlink CSI matrix in the spatial-frequency domain is expressed as:
\begin{equation} \label{eq:CSI}
\widetilde{\mathbf{H}} = \left[\tilde{\mathbf{h}}_1, \ldots, \tilde{\mathbf{h}}_{N'_{\mathrm{c}}} \right]^H \in \mathbb{C}^{N'_{\mathrm{c}} \times N_{\mathrm{t}}}.
\end{equation}

To fully exploit the advantages of multiple-antenna systems, the BS designs the precoding vectors $ \lbrace \mathbf{v}_n \mid n = 1, \ldots, N'_{\mathrm{c}} \rbrace $  based on the feedback of the downlink CSI. Given the channel's inherent sparsity, discrete Fourier transformation (DFT) is applied to transform $\tilde{\mathbf{H}}$ into the angular-delay domain:
\begin{equation} \label{eq:FHF}
\mathbf{H} = \mathbf{F}_{\mathrm{d}} \widetilde{\mathbf{H}} \mathbf{F}_{\mathrm{a}}^H,
\end{equation}
where $ \mathbf{F}_{\mathrm{d}} \in \mathbb{C}^{N'_{\mathrm{c}} \times N'_{\mathrm{c}}} $ and $ \mathbf{F}_{\mathrm{a}} \in \mathbb{C}^{N_{\mathrm{t}} \times N_{\mathrm{t}}} $ are the 2D DFT matrices in the delay and angle domains, respectively. Due to the limited channel delay spread, $ \mathbf{H} $ contains non-zero values only in the first $ N_{\mathrm{c}} $ rows, allowing the truncation of the remaining rows. Hence, we redefine $ \mathbf{H} $ as the truncated $ N_{\mathrm{c}} \times N_{\mathrm{t}} $ matrix. Although this preprocessing reduces the number of feedback parameters to $ N = 2 N_{\mathrm{c}}  N_{\mathrm{t}} $, the feedback overhead remains excessive. Thus, further compression of $ \mathbf{H} $  is essential before transmission.

\subsection{DL-Based CSI Feedback}

Conventional DL-based CSI feedback frameworks, such as CsiNet \cite{wen2018deep}, adopt an autoencoder structure, where a pre-trained NN at the UE  is responsible for CSI encoding. However, adapting this NN to varying environments often requires frequent updates at both the UE and the BS. To address this issue, one-sided CSI feedback frameworks shift all NN layers to the BS, enabling the UE to perform random linear projection for CSI compression, while the BS reconstructs the CSI using a NN \cite{guo2024onesided}. An illustration of a baseline one-sided CSI reconstruction framework is shown in Fig.~\ref{fig:framea}. The compression process is formulated as:
\begin{equation}
\mathbf{s} = \mathbf{A} \text{vec}(\mathbf{H}),
\end{equation}
where $\text{vec}(\mathbf{H})\in \mathbb{R}^{N}$ denotes the vectorization of the CSI matrix $\mathbf{H} \in \mathbb{R}^{N_{\mathrm{c}} \times N_{\mathrm{t}}}$ after separating its real and imaginary components into distinct channels, and expressing the result in a real-valued form, with $N = 2 \times N_{\mathrm{c}} \times N_{\mathrm{t}}$. The compressed codewords $\mathbf{s} \in \mathbb{R}^{M}$ satisfy $M < N$, where the compression ratio (CR) is defined as $\gamma = \frac{M}{N}$. The linear projection matrix $\mathbf{A} \in \mathbb{R}^{M \times N}$ can be randomly generated.

After compression, the codeword $\mathbf{s}$ is transmitted back to the BS, where a CSI reconstruction NN is employed to recover the original CSI matrix. Letting $\text{Rec}(\cdot)$ denote the reconstruction function, the recovered CSI matrix $\hat{\mathbf{H}}$ is given by:

\begin{equation}
\hat{\mathbf{H}} = \text{Rec}(\mathbf{s}; {\bf \Theta}_{\rm Rec}),
\end{equation}
where ${\bf \Theta}_{\rm Rec}$ represents the parameters of the CSI reconstruction NN.

The optimization objective is to minimize the mean square error (MSE) between the reconstructed CSI matrix and the original CSI matrix, defined as:
\begin{equation} \label{eq:MSE}
\mathcal{L} = \frac{1}{T} \sum_{t=1}^{T} \left\| \hat{\mathbf{H}}_t - \mathbf{H}_t \right\|^2_2,
\end{equation}
where \( T \) denotes the total number of training samples.

\section{Scene Graph-Aided DL-Based CSI Feedback}
In this section, we first present the motivation for incorporating scene graph information into DL-based CSI feedback systems. We then introduce the proposed AdapCsiNet framework, which leverages scene graph features to enable dynamic adaptation of the feedback process. Finally, we provide a detailed description of the underlying NN architecture.
\begin{figure*}[!t]
\centering
{\includegraphics[width=0.9\textwidth]{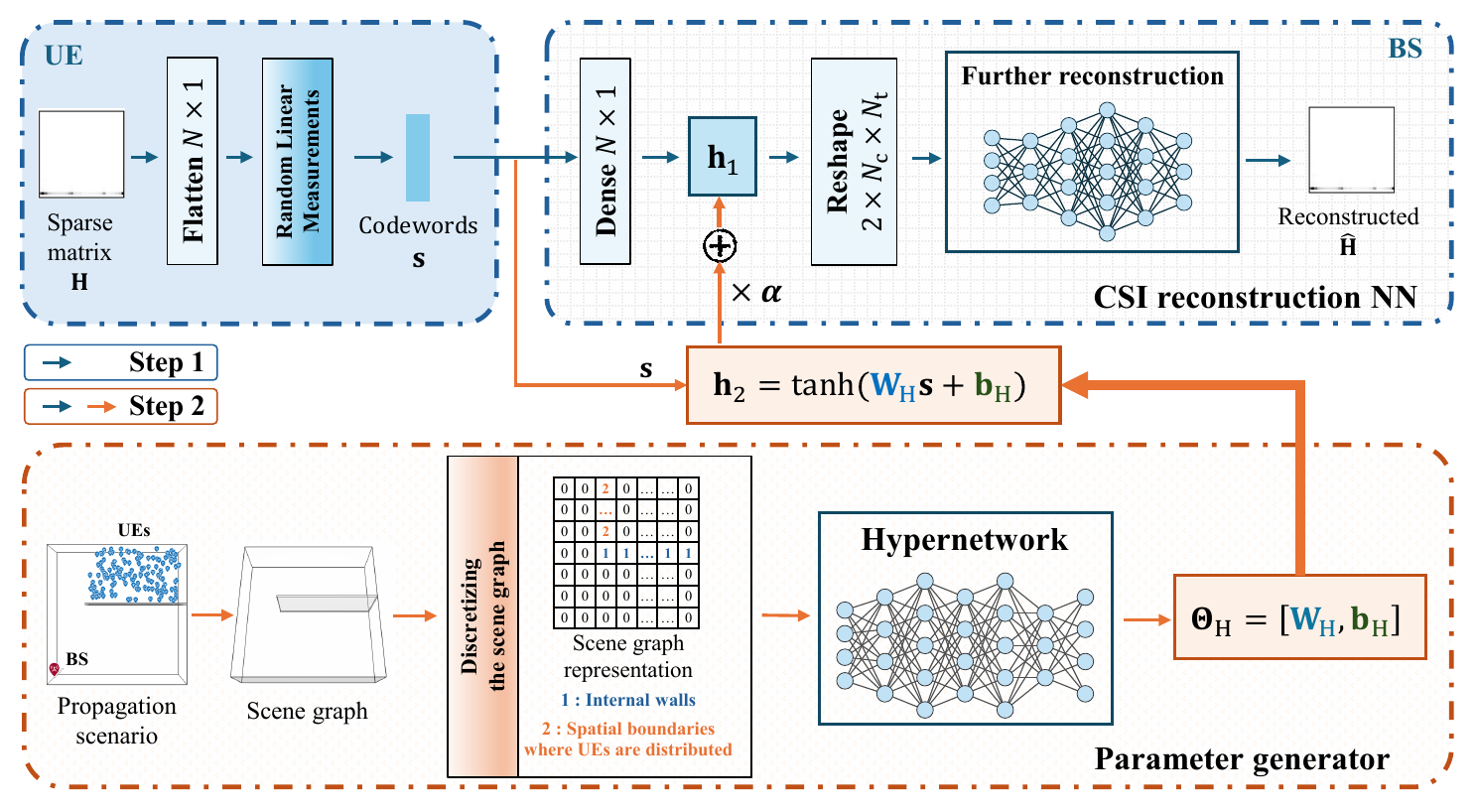}%
\caption{Illustration of the AdapCsiNet framework.}
\label{fig:frameb}}
     \vspace{-0.5cm}
\end{figure*}
 
\subsection{Motivation}
When a UE  moves into a new environment with a different spatial structure, the multipath distribution of the channel changes, leading to a significant decline in the reconstruction accuracy of a pre-trained NN. While online training can mitigate this issue by fine-tuning the NN with newly collected CSI samples, this approach imposes substantial overhead due to its high data collection and computational demands. Furthermore, treating CSI feedback purely as a data-driven task overlooks the intrinsic physical relationship between the propagation environment and channel distribution. Moreover, fine-tuned NN parameters are typically optimized for a specific scene, making adaptation to varying environments challenging.

Since the spatial structure of the communication environment, such as building layouts and obstacle distributions, directly impacts channel characteristics, integrating prior environmental knowledge into the CSI feedback framework can significantly enhance adaptability. Specifically, the representation of the physical environment provides valuable insights into channel distribution patterns, which can inform NN parameter adjustments. By establishing a structured mapping from environmental features to channel characteristics and subsequently to NN parameters, the NN can dynamically adapt to diverse propagation conditions, thereby improving CSI feedback performance.

This leads to our core idea: leveraging scene graphs to enhance environmental adaptability in CSI feedback NNs. By doing so, we eliminate the need for online CSI collection and retraining in unseen environments, thereby reducing deployment overhead while maintaining high reconstruction accuracy.

\subsection{Framework}
To enhance the environmental adaptability of CSI feedback, we propose the AdapCsiNet framework, illustrated in Fig.~\ref{fig:frameb}. The framework integrates environmental information into the CSI reconstruction process via two key components:
\begin{itemize}
    \item {\bf Parameter Generator:} Extracts environmental features from a scene graph and dynamically generates parameters for the CSI reconstruction NN using a hypernetwork.

    \item {\bf CSI Reconstruction NN:} Performs an initial CSI reconstruction via a dense layer and subsequently refines it using a residual convolutional NN.
\end{itemize}
The following subsections provide a detailed breakdown of each component.

\subsubsection{Design of Parameter Generator} The parameter generator is crucial for integrating environmental knowledge into the CSI reconstruction process. It consists of two major subcomponents:
\begin{itemize}

\item \textbf{Scene Graph Preprocessing:} 
To effectively process environmental information for input into the subsequent hypernetwork, scene graph preprocessing is performed. Fig.~\ref{fig:frameb} illustrates an example of this procedure. In this scenario, the BS position is fixed, and UEs are distributed within a region defined by an internal wall and the outer boundaries. The UE distribution region is determined by the random positions of the internal wall. The scene graph is extracted from the propagation scenario and retains essential information, including the geometric structure of the scenario, building layouts, and obstacle distributions. This scene graph is then discretized into a matrix form, where the values 1 and 2 represent, respectively, the internal wall and the UE distribution boundaries. This matrix serves as the input to the hypernetwork, enabling it to establish a relationship between spatial structures and channel characteristics.

\item \textbf{Hypernetwork:} 
AdapCsiNet employs a hypernetwork-based architecture \cite{ha2016hypernetworks} to generate environment-aware parameters for the CSI reconstruction NN. A hypernetwork is designed to generate parameters for another NN. Instead of directly learning fixed parameters, it learns a mapping function that dynamically produces parameters for a target NN, enabling efficient adaptation. By incorporating environmental information, the approach allows the NN to adapt to diverse communication environments without requiring online retraining. The hypernetwork generates environment-aware parameters for the CSI reconstruction NN as follows: 
\begin{equation}
\boldsymbol{\Theta}_{\text{H}} = \begin{bmatrix}
\mathbf{W}_{\mathrm{H}}, ~\mathbf{b}_{\mathrm{H}}
\end{bmatrix},
\end{equation}
where $\boldsymbol{\Theta}_{\text{H}} \in \mathbb{R}^{N \times (M+1)}$ represents the complete parameters produced by the hypernetwork, consisting of the weight matrix $\mathbf{W}_{\mathrm{H}} \in \mathbb{R}^{N \times M}$ and the bias vector $\mathbf{b}_{\mathrm{H}} \in \mathbb{R}^{N \times 1}$. These parameters capture the environmental characteristics, enabling the CSI reconstruction NN to adaptively reconstruct CSI in diverse environments.  
\end{itemize}

After processing with the two subcomponents described above, the parameter generator produces the environment-aware parameters $\boldsymbol{\Theta}_{\text{H}}$, which are then used in conjunction with the compressed codeword $\mathbf{s}$ to obtain an enhanced CSI estimate. Specifically, the estimate $\mathbf{h}2 \in \mathbb{R}^{N}$ is computed from the compressed codeword $\mathbf{s} \in \mathbb{R}^{M}$ as: 

\begin{equation}
\mathbf{h}_2 = \tanh(\mathbf{W}_{\mathrm{H}} \mathbf{s} + \mathbf{b}_{\mathrm{H}}),
\end{equation}
where a non-linear activation function $\tanh(\cdot)$ is applied to better capture the complex mapping between scene information and CSI distribution.

\subsubsection{Design of CSI Reconstruction NN}
As shown in Fig.~\ref{fig:framea}, conventional CSI feedback typically relies on a one-sided CSI reconstruction NN. In AdapCsiNet, this structure is retained but enhanced with the integration of environmental information, as depicted in Fig.~\ref{fig:frameb}. The CSI reconstruction NN comprises two main components: 
\begin{itemize}
\item \textbf{Initial Reconstruction (Dense Layer):} The first stage of reconstruction is performed by a dense layer that takes the compressed CSI codeword $\mathbf{s}$ and produces an initial CSI estimate $\mathbf{h}1 \in \mathbb{R}^{N}$. Since this dense layer contains the largest number of parameters in the reconstruction process, it significantly influences reconstruction accuracy. To improve adaptability, the dense layer parameters are fine-tuned using the outputs generated by the hypernetwork. To balance adaptation with accuracy preservation, a regularization parameter $\alpha < 1$ is introduced, ensuring that the refined CSI estimate remains consistent with the initial reconstruction. By adjusting $\alpha$, the contribution of the scene graph information to the CSI reconstruction can be controlled. The final refined CSI input for further reconstruction is computed as:
\begin{equation} 
\mathbf{h}_{\text{input}} =  \mathbf{h}_1 + \alpha \mathbf{h}_2.
\end{equation}

\item \textbf{Further Reconstruction (Residual Convolutional NN):} The refined estimate $\mathbf{h}_{\text{input}}$ is first reshaped into $\mathbf{H}_{\text{input}} \in \mathbb{R}^{2 \times N_{\mathrm{c}} \times N_{\mathrm{t}}}$, which is then processed by a residual convolutional NN. This NN refines and denoises the initial estimate by learning from the residuals between the predicted CSI and the true CSI values, thereby improving reconstruction accuracy. The final output, $\mathbf{\hat{H}}$, represents the enhanced CSI estimate that effectively incorporates environmental awareness.  
\end{itemize}

\subsubsection{Training Process}
To ensure robust baseline performance of the one-sided CSI reconstruction NN while enabling efficient environment-aware adaptation, AdapCsiNet is trained in a two-step process.
\begin{itemize}
    \item \textbf{Step 1---Pre-training of  One-sided CSI Reconstruction:} As illustrated by the blue arrow in Fig.~\ref{fig:frameb}, this step involves pre-training a generalized CSI feedback NN using mixed CSI samples collected from multiple environments. During pre-training, the NN takes the original CSI matrix $\mathbf{H}$ as input and outputs the reconstructed matrix $\mathbf{\hat{H}}$. The training objective is to minimize the MSE between $\mathbf{H}$ and $\mathbf{\hat{H}}$, defined as:
    \begin{equation}  
        \mathcal{L} = \mathbb{E} \left\{ \|\mathbf{H} - f_{\mathrm{Step1}}(\mathbf{H})\|_2^2 \right\},  
    \end{equation}  
where $f_{\mathrm{Step1}}(\mathbf{H})$ represents the reconstructed matrix $\mathbf{\hat{H}}$ obtained through the pre-trained baseline one-sided CSI feedback NN, and \(\|\cdot\|_2\) denotes the Euclidean norm.

    \item \textbf{Step 2---Refinement with Environment Information via Hypernetwork:}  
    In Step 2, the hypernetwork-based parameter generator is introduced to refine the pre-trained CSI reconstruction NN by incorporating scene graph information. Specifically, the framework takes two inputs: the original CSI matrix $\mathbf{H}$ for the CSI reconstruction NN and the scene graph representation matrix for the hypernetwork. The hypernetwork generates environment-aware parameters, enabling the refined reconstruction $\mathbf{\hat{H}}$ to adapt to diverse propagation environments. The training objective remains minimizing the MSE between $\mathbf{H}$ and the refined reconstruction, defined as:
\begin{equation}  
        \mathcal{L}' = \mathbb{E} \left\{ \|\mathbf{H} - f_{\mathrm{Step2}}(\mathbf{H}; \boldsymbol{\Theta}_{\text{H}}\|_2^2 \right\},  
    \end{equation} 
where $\boldsymbol{\Theta}_{\text{H}}$ represents the environment-aware parameters derived by the hypernetwork from the scene graph, and $f_{\mathrm{Step2}}(\mathbf{H}; \boldsymbol{\Theta}_{\text{H}})$ denotes the refined final reconstruction $\mathbf{\hat{H}}$ that incorporates environmental information.

\end{itemize}

\begin{figure}[!t]
\centering
\subfloat[]{
  \includegraphics[width=1.4in]{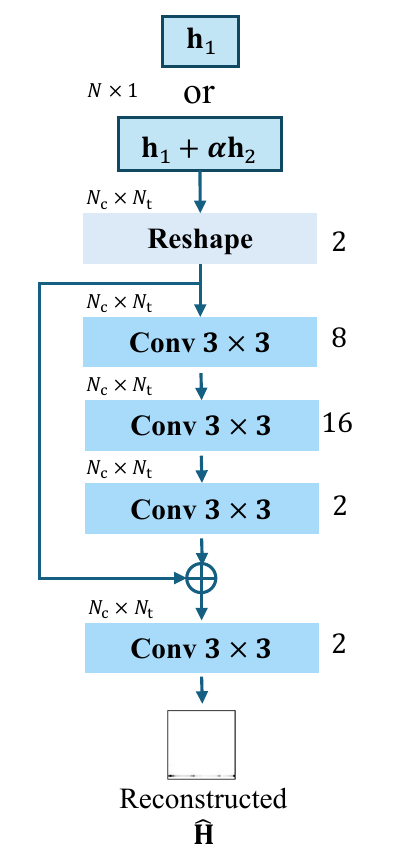}
  \label{fig:refine}  
}
\hfill
\subfloat[]{
  \includegraphics[width=1.4in]{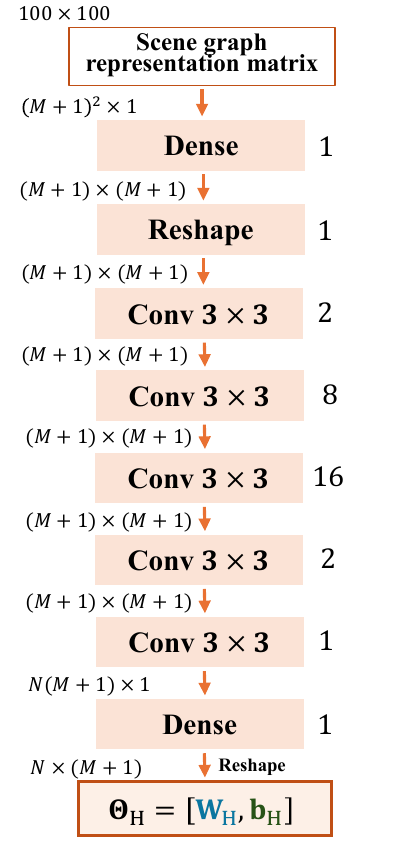}
  \label{fig:hyper} 
}
\caption{Detailed NN architecture. (a) CSI reconstruction NN. (b) Hypernetwork structure.}
\label{fig:NNdetailed}
\vspace{-0.5cm}
\end{figure}

\subsection{Detailed NN Architecture}

\begin{itemize}[leftmargin=*]  
\item \textbf{CSI reconstruction NN:} As illustrated in Fig.~\ref{fig:NNdetailed}\subref{fig:refine}, the CSI reconstruction NN first processes the initially reconstructed matrix \( \mathbf{h}_1 \) or the refined estimate \( \mathbf{h}_1 + \alpha \mathbf{h}_2 \). This estimate is then reshaped into a matrix of dimensions $2 \times N_{\mathrm{c}} \times N_{\mathrm{t}}$ before being fed into three sequential convolutional layers with $3 \times 3$ kernels (featuring 8, 16, and 2 channels). Finally, a dedicated convolutional layer outputs the final reconstructed CSI \( \mathbf{\hat{H}} \).

\item \textbf{Hypernetwork:} 
As shown in Fig.~\ref{fig:NNdetailed}\subref{fig:hyper}, the hypernetwork first applies a dense layer to process the scene graph representation matrix and extract feature information. The extracted features are then refined through five convolutional layers before being mapped to the parameter space by a final dense layer, which produces the environment-aware parameters $\boldsymbol{\Theta}_{\text{H}}$. These parameters are subsequently used to fine-tune the pre-trained CSI reconstruction NN, thereby enhancing its adaptability to different environments.

\end{itemize}

\section{Simulation results}\label{sec:performance}
This section presents the channel generation process and hyperparameter settings, followed by an evaluation of the performance gains achieved by the proposed method.

\subsection{Simulation Setting}\label{subsec:simulation}

\subsubsection{Dataset Generation}

To validate the proposed method, it is necessary to generate scene graphs and corresponding CSI datasets simultaneously, ensuring that the scene graphs accurately reflect the channel distributions. Using real-world CSI data would require constructing physical scenes, deploying hardware, and extensive data collection, which is both costly and labor intensive. Alternatively, we adopt a ray tracing approach---a widely used technique for channel modeling. Ray tracing simulates multipath channels in detailed 3D environments, accurately representing propagation conditions by accounting for reflection, diffraction, and scattering. The extracted parameters are then used to synthesize high-fidelity CSI datasets, and the proposed method employs these datasets while simultaneously obtaining the corresponding scene graphs.

Specifically, indoor environments were modeled using a graphics software tool, Blender\footnote{https://www.blender.org/}, a popular tool for creating 3D indoor scene models. The constructed environments were processed using a ray tracing simulator implemented in MATLAB\footnote{https://www.mathworks.com/help/comm/ref/rfprop.raytracing.html} to generate channel characteristics. For channel generation, the maximum number of reflections was set to 2, and the maximum number of diffractions to 1. Table \ref{tab:rtparams} lists the complete set of parameters used in the ray tracing and channel generation process.

\begin{table}[htbp]
\centering
\caption{Channel Generation Parameters}
\label{tab:rtparams}
\begin{tabular}{|l|l|}
\hline
\textbf{Parameter}           & \textbf{Value}         \\ \hline
Array number                 & Tx 8, Rx 1                   \\ \hline
Subcarriers                  & 256                    \\ \hline
Center frequency             & 5.8 GHz                \\ \hline
Bandwidth                    & 20 MHz                 \\ \hline
Indoor area                  & 10 m × 10 m × 3 m         \\ \hline
BS height                    & 2.9 m                  \\ \hline
UE height                  & 0.8 m                  \\ \hline
MaxNumReflections            & 2                      \\ \hline
MaxNumDiffractions           & 1                      \\ \hline
\end{tabular}
\end{table}

\begin{figure}[!t]
\centering
\subfloat[]{
  \includegraphics[width=1.52in]{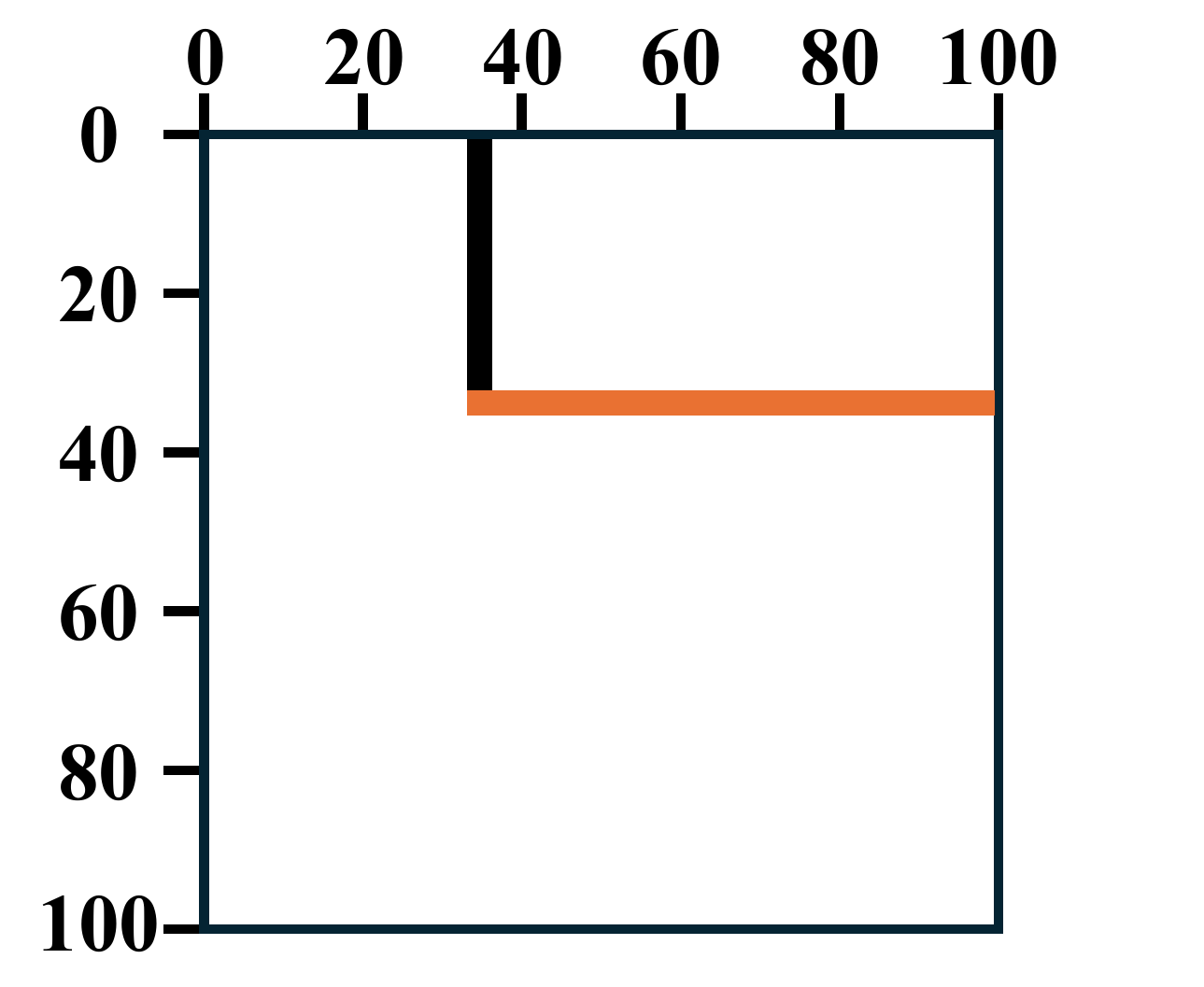}
  \label{fig:matrix}   
  
}
\hfill
\subfloat[]{
  \includegraphics[width=1.65in]{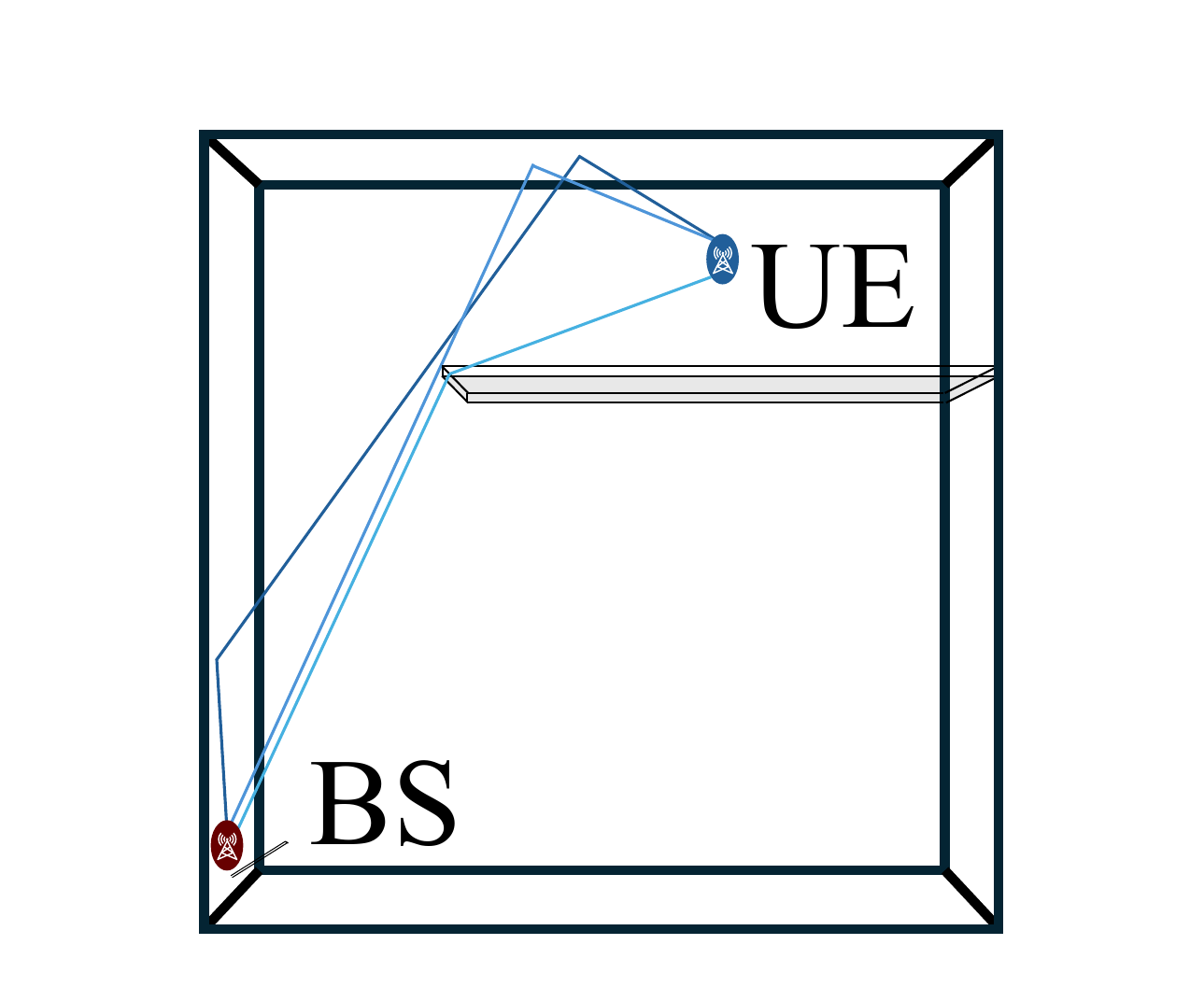}
  \label{fig:rtpath} 
} 
\caption{Visualization of the scene graph. (a) Discrete scene graph matrix \(100 \times 100\) where 1 (orange) denotes internal walls, 0 (blank areas) represents free space, and 2 (black) indicates spatial boundaries where UEs are distributed. (b) Ray tracing path visualization of a UE in the scene.}
\label{fig:comparison}
\vspace{-0.5 cm}
\end{figure}

Using this methodology, 200 different indoor scene models were created in Blender. While all scenes share the same overall dimensions, the length and distribution of internal walls were randomized. The wall material was assumed to be wood. To emphasize the impact of the propagation environment on CSI distribution, UEs were randomly placed within an enclosed space defined by both internal and outer walls. The BS position, as shown in Fig.~\ref{fig:comparison}\subref{fig:rtpath}, remained fixed across all scenes. 
The corresponding scene graph representation was discretized into  a $100 \times 100$ matrix format, as illustrated in Fig.~\ref{fig:comparison}\subref{fig:matrix}. In this matrix, internal walls are denoted by 1 (orange), blank areas (free space) are denoted by 0, and boundaries where UEs are distributed are denoted by 2 (black). This scene graph matrix serves as input to the hypernetwork, encoding environmental structure information for adaptation.
Fig.~\ref{fig:comparison}\subref{fig:rtpath} shows the ray tracing propagation paths of a UE within the scene, demonstrating how the simulated environment affects CSI.

To ensure a robust evaluation, 1,000 CSI samples were randomly selected from each environment. The dataset was then partitioned into three subsets: 160 environments (corresponding to 160,000 samples) were allocated for training, while 20 environments (comprising 20,000 samples) were designated for validation. The remaining 20 environments (also 20,000 samples) were reserved for testing. Notably, the test environments were entirely unseen during training, ensuring that the evaluation accurately reflects the model’s generalization capability across novel environments.

In Step 2, the pre-trained model is fine-tuned using environmental information. The dataset partitioning remains the same, but each sample is now paired with its corresponding scene graph. This scene graph information is fed into the hypernetwork, enabling the CSI feedback NN to adapt to different environments dynamically.

\subsubsection{Hyperparameter Setting}
All training and testing were conducted on an NVIDIA DGX-1 platform equipped with an NVIDIA Tesla V100 GPU, using the TensorFlow library for model implementation. The NN parameters were initialized with the Glorot uniform initializer, and optimization was performed using the Adam optimizer.

In Step 1, which involves the pre-training of the one-sided CSI reconstruction NN, the batch size was set to 200, and the model was trained for 1,000 epochs with a learning rate of 0.001. In Step 2, where the NN is refined using environmental information via the hypernetwork, the batch size, number of epochs, and initial learning rate remained the same as in Step 1. To enhance training efficiency, an adaptive learning rate strategy was employed, where the learning rate was halved if the validation loss did not improve within 30 epochs.

A crucial aspect of Step 2 refinement is the integration of environmental information through the scene graph. To balance the influence of this additional information while maintaining the accuracy of CSI reconstruction, the regularization parameter \( \alpha \) for the fully connected layers was set to 0.6. Experimental results indicate that this choice of \( \alpha \) prevents the scene matrix information from excessively dominating the CSI reconstruction process while still effectively utilizing environmental knowledge to improve reconstruction accuracy.

\subsubsection{Performance Metrics}
To evaluate the accuracy of CSI reconstruction, the normalized mean squared error (NMSE) is adopted, as it provides a standardized measure for comparing different reconstruction methods. The NMSE is defined as 
\begin{equation}  
\text{NMSE} = \frac{1}{T} \sum_{i=1}^{T} \frac{\|\hat{\mathbf{H}}^{[i]} - \mathbf{H}^{[i]}\|_2^2}{\|\mathbf{H}^{[i]}\|_2^2}
\end{equation}
where \( \hat{\mathbf{H}}^{[i]} \) represents the estimated CSI matrix for the \( i \)-th sample, while  \( \mathbf{H}^{[i]} \) is the corresponding original CSI matrix. The parameter \( T \) denotes the total number of training samples.

\subsection{Result Analysis}
The primary objective of the proposed scheme is to validate the effectiveness of leveraging environmental information to enhance CSI reconstruction. To achieve this, we compare the proposed scene graph-aided DL-based CSI feedback scheme (AdapCsiNet) with a general feedback NN. The latter refers to a NN pre-trained using mixed CSI data from multiple environments \cite[Fig. 4]{mix2024train} in Step 1 without incorporating scene graph features. This comparison highlights the performance gains achieved by integrating scene graph information into the reconstruction process.

\begin{figure}[t!]
\centering
\includegraphics[width=2.8in]{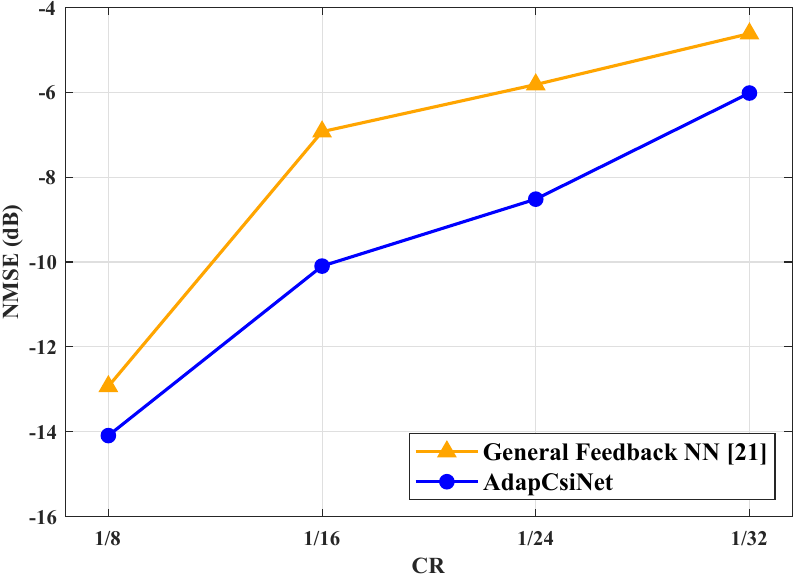} 
\caption{Comparison of NMSE under different CRs for the general feedback NN \cite{mix2024train} and the proposed AdapCsiNet.}
\label{nmse_comparison}
\end{figure}

Figure~\ref{nmse_comparison} presents the NMSE performance of CSI reconstruction for both schemes under different compression ratios (CRs). The CRs are set to 1/8, 1/16, 1/24, and 1/32. As shown in the figure, introducing scene graph features significantly enhances the CSI reconstruction performance compared to the general feedback NN \cite{mix2024train}, particularly at lower CRs. Notably, at a CR of 1/24, AdapCsiNet achieves the highest performance gain, improving the NMSE by 46.4\% (2.7 dB) over the general feedback NN \cite{mix2024train}. These results demonstrate that environmental information extracted from scene graphs effectively assists the CSI reconstruction process by providing valuable insights into the underlying propagation characteristics of the communication environment. 

\begin{figure}[t]
\centering
\includegraphics[width=2.8in]{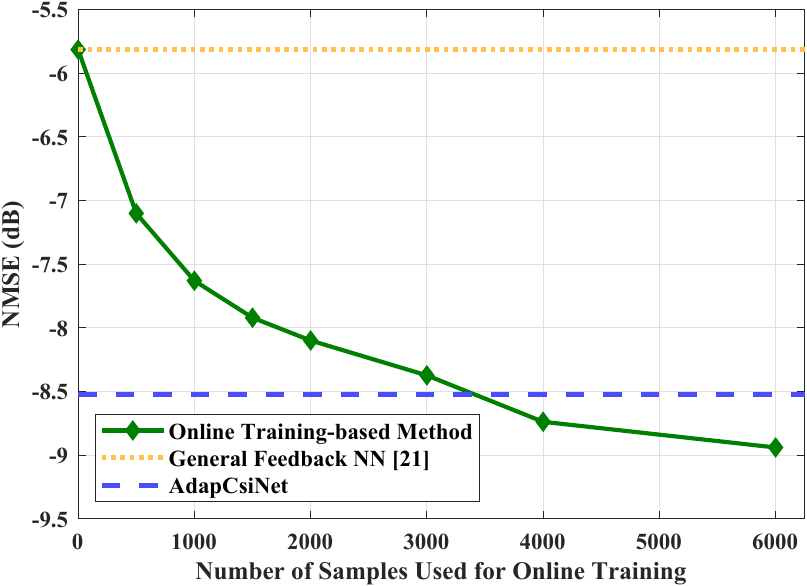}
\caption{Impact of the number of CSI samples in the online training method at CR = 1/24.}
\label{online_performance}
\vspace{-0.5cm}
\end{figure}

Further analysis is conducted to examine the training overhead associated with the proposed scene-graph-assisted CSI feedback scheme compared to the online training scheme. The results, illustrated in Fig.~\ref{online_performance}, depict the relationship between NMSE at CR = 1/24 and the number of fine-tuned CSI samples used in online training. In the figure, the orange dashed line represents the NMSE of the general feedback NN \cite{mix2024train}, which does not incorporate scene graph features, while the blue dashed line represents the NMSE of the scene-graph-assisted CSI feedback scheme, which achieves -8.52 dB. The green solid line indicates the NMSE of the online training scheme with varying numbers of training samples.

From Fig.~\ref{online_performance}, it is evident that as the number of CSI samples increases, the reconstruction accuracy of the online training scheme gradually improves. This suggests that training a scene-specific NN using an adequate number of samples can lead to better reconstruction performance. However, when the number of training samples exceeds 4,000, additional samples provide only marginal improvements in feedback performance. Furthermore, when the number of training samples reaches 3,000, the performance of the online training scheme becomes nearly identical to that of AdapCsiNet. These findings suggest that AdapCsiNet can achieve comparable performance without requiring extensive CSI data collection, unlike the online training scheme, which relies on substantial data collection efforts for fine-tuning.

\begin{figure}[t]
\centering
\includegraphics[width=2.8in]{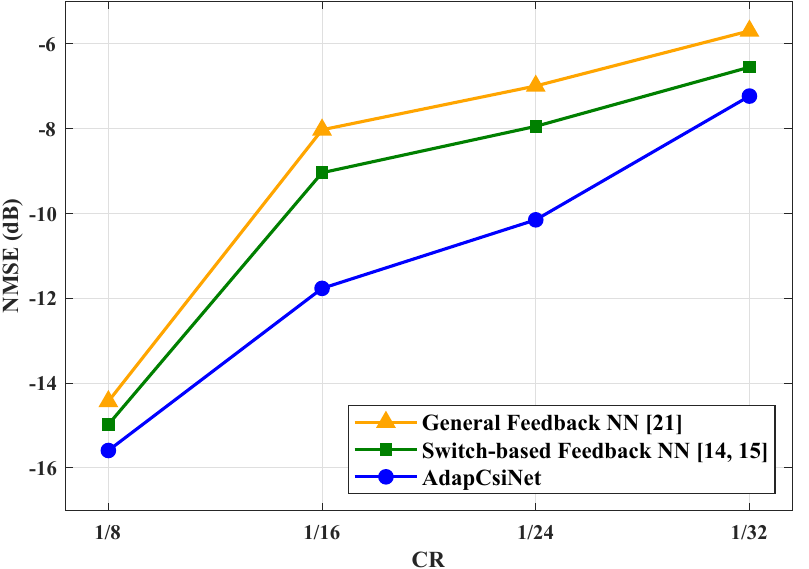}
\caption{Comparison of CSI reconstruction NMSE between AdapCsiNet and the switch-based approach \cite{zhang2023switch,bathala2023multi} in LOS environments, tested on LOS-specific CSI samples.}
\label{LOS_comparison}
\vspace{-0.5cm}
\end{figure}

To further highlight the distinction between AdapCsiNet and the switch-based approach \cite{zhang2023switch,bathala2023multi}, environments are categorized into two types: line-of-sight (LOS)-dominated environments and non-line-of-sight (NLOS)-dominated environments. A dedicated LOS-specific NN is trained based on the general feedback NN structure \cite{mix2024train}, and its performance is compared against AdapCsiNet. The results, shown in Fig.~\ref{LOS_comparison}, are obtained by testing all methods exclusively on LOS-specific CSI samples. Despite being tested in a specialized LOS environment, AdapCsiNet still outperforms the switch-based method, demonstrating its superior adaptability.

The performance improvement achieved by AdapCsiNet can be attributed to its ability to extract fine-grained environmental features from scene graphs, allowing it to dynamically generate customized NN parameters for each environment. In contrast, the switch-based method relies on coarse-grained information to determine the appropriate NN, limiting its ability to adapt to subtle variations within the environment. These results confirm that extracting scene-specific features from scene graphs significantly enhances CSI reconstruction performance, making AdapCsiNet a more effective and efficient solution for multi-environment CSI feedback.

\section{conclusion}
This study proposes AdapCsiNet, a deep learning-based CSI feedback framework that leverages environmental features to enhance reconstruction accuracy. By integrating scene graph information through a hypernetwork structure, AdapCsiNet enables NNs to effectively capture and utilize the mapping between environmental characteristics and CSI distribution. Simulation results demonstrate that the proposed framework fully exploits environmental knowledge, significantly improving CSI reconstruction performance compared to conventional methods. Moreover, when compared with typical online training approaches, AdapCsiNet achieves comparable feedback accuracy without requiring extensive CSI data collection, thereby reducing system overhead and enhancing deployment efficiency. These findings underscore the potential of environment-aware deep learning in advancing CSI feedback mechanisms for next-generation wireless communication systems.
\vspace{-0.5cm}

\bibliographystyle{IEEEtran}
\bibliography{reference.bib}

\end{document}